\documentstyle[prl,twocolumn,aps,epsf]{revtex}
\begin{document}

\twocolumn[
\hsize\textwidth\columnwidth\hsize\csname@twocolumnfalse\endcsname
\draft

\title{Temperature dependent resistivity of spin-split subbands
in GaAs 2D hole system}
\author{E. H. Hwang and S. Das Sarma}
\address{Condensed Matter Theory Center,
Department of Physics, University of Maryland, College Park,
Maryland  20742-4111 } 
\date{\today}
\maketitle

\begin{abstract}
We calculate the temperature dependent resistivity in spin-split
subbands induced by the inversion asymmetry of the confining 
potential in GaAs 2D hole systems. 
By considering both temperature dependent multisubband screening
of impurity disorder and hole-hole scattering
we find that the strength of the metallic behavior depends on 
the symmetry of the confining potential (i.e., spin-splitting) over a
large range of hole density. At low density above the metal-insulator
transition  we find that effective disorder 
reduces the enhancement of the 
metallic behavior induced by spin-splitting.
Our theory is in good qualitative agreement with existing experiments.


\end{abstract}
\vspace{0.3in}
]

\newpage

The observation of an apparent ``metallic'' behavior in low density 
two dimensional (2D) electron or hole semiconductor systems has attracted 
a great deal of recent attention \cite{one} because of the long-standing 
belief that all 2D (disordered) electronic systems (at least in the 
noninteracting limit) are insulators (i.e., infinite zero temperature 
resistance in the thermodynamic limit). Although the experimental 
situation is controversial (to say the least) because the ``metallicity'' 
of the 2D system must invariably be inferred from an extrapolation 
(to zero temperature) of finite temperature (usually $T \ge 100 mK$) 
data, there is broad agreement on a number of intriguing and manifestly 
interesting features of the observed 2D resistivity, $\rho(T,n)$, as a 
function of temperature ($T$) and 2D free carrier density ($n$), that 
require theoretical understanding irrespective of whether there is a 
true $T=0$ 2D metal or not. 


One of the most striking experimental anomalies in this subject is the 
strong temperature dependence of the resistivity of this effective 2D 
conducting phase at low carrier densities and in the relatively modest 
temperature range of $0.1K - 4K$ where the measured resistivity may 
change by as much as a factor of three whereas the corresponding 
resistivity change in this temperature range (the so-called 
Bloch-Gr\"{u}neisen behavior) for a real 3D metal (e.g., Cu, Al) is 
miniscule (less than $1\%$). In fact, very weak temperature dependence 
of resistivity is also observed in 2D electron systems at higher densities, 
but at lower densities (and in relatively clean samples) $\rho(T)$ shows 
strikingly strong temperature dependence, not found in any known metallic 
systems. This very strong dependence of $\rho(T,n)$ on $T$ at low $n$, 
specifically in 2D p-GaAs hole systems, is the subject matter of this paper.

There have been two alternate concrete physical mechanisms proposed in 
the literature to explain the strong temperature dependence of $\rho(T,n)$ 
in the 2D ``metallic'' systems. One mechanism \cite{hwang} is the possible 
temperature dependence of effective impurity disorder in the system, 
arising, for example, from the temperature dependence of finite wave 
vector 2D screening which could be very strong at low carrier densities. 
This mechanism \cite{hwang,stern,sds} is quite universal in the sense that 
all 2D carrier systems should manifest some aspects of temperature 
dependent screening since random (unintentional) charged impurities 
(and often, particularly in modulation doped 2D systems, remote dopants 
as well) are invariably present in semiconductors. We have earlier shown 
through detailed transport calculations\cite{hwang} 
that the temperature dependent 
disorder scattering could in fact provide a reasonable semi-quantitative 
explanation for the observed temperature dependence of 2D resistivity 
behavior in n-Si MOSFETs, p-SiGe, and p-GaAs 2D systems. The second 
mechanism for the temperature dependence of resistivity, which applies 
essentially only to 2D p-GaAs hole systems in a quantitatively 
significant manner, has been discussed extensively by the Princeton 
group and others \cite{papa,ham,murzin}. 
This manifestly non-universal mechanism involves a 
temperature dependent transport mechanism arising from intersubband hole 
scattering between two spin-split hole bands, which are present in the 
GaAs valence band. The magnitude of this effect should therefore 
correlate with the valence band spin-splitting as has been demonstrated 
experimentally. This proposal has been somewhat controversial, and there 
are experimental claims (and counter-claims) for nonexistence (and the 
existence) of this spin-splitting induced intersubband scattering 
mechanism. It is obvious that both of these physical mechanisms must 
be simultaneously operational in the p-GaAs system because scattering 
by random charged impurities and valence band spin-splitting 
are both unavoidably present in real 
2D GaAs hole structures. What is therefore required is a (technically 
highly demanding) theory which includes both mechanisms on an 
equal footing, and in this paper we provide such a theory which, as we 
describe below, qualitatively explains all of the observed features of 
the temperature dependent $\rho(T,n)$ of 2D GaAs hole systems. Our work 
convincingly shows that in order to understand the 2D GaAs hole data 
comprehensively a theory such as ours including both mechanisms 
equivalently is necessary although one mechanism or the other may very 
well be dominant in specific experimental p-GaAs systems depending on 
the details of disorder and spin-splitting (which would vary from sample 
to sample), thus resolving the existing controversy of whether 
spin-splitting is a relevant mechanism in 2D ``metallic'' behavior or 
not. In all our discussions in this paper, consistent with the accepted 
(but, not quite rigorous) terminology in this subject, the word 
2D ``metal'' or ``metallic'' behavior will only mean a strong temperature 
dependence of $\rho(T)$ at a fixed density (above the so-called 
metal-insulator transition ``critical'' density), with $d\rho/dT$ being 
positive at low temperatures.

The valence band holes in GaAs-AlGaAs heterostructures have large 
spin-orbit interaction due to
the inversion asymmetry of the zincblende 
crystal structure and the real space asymmetry induced by the 
confining potential, producing
substantial zero-magnetic-field spin-splitting \cite{eisen}.
The lack of inversion symmetry in the p-GaAs system
can lift the spin degeneracy 
and produce two groups of spin-split 2D subbands of holes with different 
band dispersions and transport properties.
Recently, Papadakis {\it et al.} \cite{papa} have experimentally
studied the relationship between the 2D metallic behavior
and the spin-splitting
of the p-GaAS quantum well systems by tuning the symmetry of the well
confinement potential while keeping the total hole density constant.
They find that over a large range of densities
(in a regime where the samples exhibit metallic behavior)
the symmetry of the quantum well plays an important role
in the sense that the temperature dependence of $\rho(T)$ seems to correlate
with the (well asymmetry tuned) spin-splitting.
At high hole densities, far above the critical density 
for the apparent metal-insulator transition (MIT), 
the metallic behavior 
is very pronounced as the spin-splitting becomes large, 
and the two spin-split subbands are occupied with
unequal densities.
This enhancement of the metallic behavior is 
induced by the temperature dependent intersubband
hole-hole scattering because
at constant total density the intersubband scattering 
becomes stronger as the spin-splitting increases.
Thus, the temperature dependent hole-hole scattering may, by itself,
give rise to the apparent metallic behavior in the spin-split
situation, at least for large carrier densities.
In low density samples \cite{papa,ham}, however,
the spin-splitting becomes much smaller, and
the change of the resistivity with temperature arising from 
intersubband hole-hole scattering is suppressed even in the presence 
of large asymmetry in the confining potential.
Strong metallic behavior of $\rho(T)$ at low densities 
cannot therefore be explained solely by spin-splitting effects. 
This indicates that additional 
scattering mechanisms other than hole-hole scattering are required 
to explain the metallic behavior at low densities.
Hamilton {\it et al.}\cite{ham} claim, based on the interpretation 
of their own 2D GaAs hole data,  that in the low density systems 
the metallic behavior is determined not by
the spin-splitting induced intersubband carrier-carrier scattering,
but by the magnitude of the 
low temperature resistivity, that is, by the effective 
disorder in the system.

In this paper we investigate the 2D hole metallic behavior observed in p-type 
GaAs systems including both disorder and spin-splitting effects.
This system forms two spin-split hole subbands (without any band offset)
with  very different effective masses.
Our goal is to calculate the temperature 
dependent resistivity in the spin-split two subband system.
We use the Boltzmann transport equation approach to calculate 
the low temperature ohmic resistivity of the two subband system  
taking into account the single particle relaxation 
times of each subband (as arising from charged impurity scattering),
$\tau_i$, and the hole-hole intersubband scattering between the
two groups of holes, $\tau_{hh}$, which relaxes the relative momentum
of the two carriers \cite{gant,appel} and therefore contributes to 
the net resistivity of the two subband system.
We calculate the single particle relaxation times $\tau_i$ 
from a multisubband transport theory taking into account
the long range Coulombic scattering by random static 
charged impurity centers invariably present in semiconductor structures.
Since screening plays a crucial role in the temperature dependence
of resistivity \cite{hwang} we consider both hole-impurity 
and hole-hole Coulomb interactions being screened on an equal footing
by the 2D hole gas in the random phase approximation (RPA) using a 
multisubband screening formalism.
In the multisubband system
screening is enhanced since all subbands contribute.
The screening effect can, in the RPA, be included in terms of a matrix 
dielectric formalism \cite{ando,siggia},
$\epsilon_{ij,i'j'}(q) = \delta_{ij}\delta_{i'j'} - 
v(q) \Pi_{ij}(q)F_{ij,i'j'}(q)$, 
where $i,j$ are the band indices, $v(q)=2\pi e^2/(\kappa q)$ with
the background lattice dielectric constant of the sample $\kappa$, 
$F_{ij,i'j'}(q)$ is the subband form factor determined by the quantum 
confinement potential, and $\Pi_{ij}(q)$ are the polarizabilities
corresponding to the two spin-split ($i,j=1,2$) subbands.
Since suppression of screening by disorder scattering may be important 
(particularly at low densities)
we include collisional broadening in the screening function
through the Dingle temperature approximation\cite{sds}.

We assume that the spin-split subbands have parabolic 
energy bands
with spin degeneracy $g_s=1$ (in each band) and isotropic Fermi surfaces
with effective masses $m_1$ and $m_2$. 
In our model the hole population of each subband,
for a fixed total density $n=n_1+n_2$,  is determined by the effective mass
ratio, i.e., $n_1/n_2 = m_1/m_2$, where $m_i$, $n_i$ are the effective 
mass and population density for each group of carriers.
Since there are two groups of holes we need to consider 
intersubband hole-hole scattering, which contributes to resistivity 
in addition to hole scattering by charged impurities.
In particular, hole-hole scattering plays an important role when the
carriers in each band have different masses and densities.
In a two-subband system
the resistivity without intersubband hole-hole scattering
is given by $\rho(T) = 1/\left [ \sigma_1(T) + \sigma_2(T) \right ]$,
where $\sigma_i=n_ie^2\langle\tau_i\rangle/m_i$ is the individual 
conductivity of the $i$th subband.
These $\tau_i$ can be calculated by using the usual Boltzmann transport
equation, which becomes a set of coupled transport
equations for distribution functions
associated with each subband\cite{siggia}.

The resistivity in the presence of 
hole-hole intersubband scattering
of the two-band system is given by \cite{gant,appel}
\begin{equation}
\rho(T) = \frac{\frac{n_2}{n_1}\sigma_1 + \frac{n_1}{n_2}\sigma_2 +\sigma}
{\frac{n^2}{n_1n_2}\sigma_1\sigma_2 + (\sigma_1+\sigma_2)\sigma},
\label{rho2}
\end{equation} 
where $\sigma_i$ is the conductivity associated 
with the $i$th subband with the scattering times 
$\tau_i$,  and 
$\sigma = e^2 n \tau_{hh} m/m_1m_2$ is the conductivity 
associated solely with hole-hole
scattering with an average mass given by 
$m=(m_1 n_1 + m_2 n_2)/n$ and a hole-hole
relaxation time ($\tau_{hh}$) for the relative momentum 
of the two spin-split carrier systems.
We now note an important feature of our spin-split two subband system: 
Spin in each subband is conserved and therefore, Coulomb scattering, 
being spin 
independent, cannot cause {\it real} intersubband scattering 
(which would change the spin index of the scattered carrier).
Thus, only {\it intra-subband} transitions contribute to the 
intersubband hole-hole relaxation time, and we find
\begin{equation}
\frac{1}{\tau_{hh}} = \frac{2g_s^4(k_BT)^2}{3 (2\pi)^2} mn 
\frac{m_1m_2}{n_1n_2}I(p),
\label{tauhh}
\end{equation}
where $g_s=1$ is the subband  degeneracy and
$I(p) = p/(2\pi)\int_{0}^{\pi}d\theta \sin\theta  W(q) 
[f(p)]^2$,
where $f(p) = (1 + p^2 + 2p\cos\theta)^{1/2}$ with $p=(n_1/n_2)^{1/2}$ 
and $q=(4\pi/g_s)\sqrt{n_1n_2}\sin(\theta)/f(p)$,
and the collisional probability 
$W(q) = 2\pi \left | {v(q)}/{\varepsilon(q)}\right |^2 $.
In long wavelength Thomas-Fermi screening 
approximation (TFA) we have
$W_{TFA}(q) = (2\pi)^3/[g_s (m_1+m_2)]^2$ in our 2D system. 
In the absence of any real 
hole transitions from one subband to the other the RPA 
dielectric function is given by
$\epsilon(q)= 1 -  v(q)[\Pi_{11}(q)+\Pi_{22}(q)]$.
The hole-hole scattering time $\tau_{hh}$ is symmetric with respect 
to the interchange of indices 1 and 2.
Note that $\tau_{hh}^{-1} \propto T^2$ while 
$\tau_i^{-1} \propto T$ for $T/T_F \ll 1$ but the two asymptotic regimes 
need not coincide (additionally, the Dingle temperature suppresses 
temperature dependence for $T \ll T_D$), and therefore 
the net 
temperature dependence could be quite complex.
When $\tau_{hh} \gg \tau_i$, which happens at very low temperatures, 
the total conductivity becomes 
the sum of the conductivities of each subband. 
When hole-hole scattering is very strong (or impurity scattering 
very weak), $\tau_{hh} \ll \tau_i$,  
Eq. (\ref{rho2}) become
$\rho(T) = (n_2^2\sigma_1 + n_1^2 \sigma_1)/(n^2 \sigma_1 \sigma_2)$.
Thus, in these two limits $\tau_{hh}$ makes  negligible
contribution to the system resistivity. The resistivity 
has a linear temperature dependence in these limits
(provided $T \ll T_F$, and Dingle temperature effects are negligible). 
When $\tau_{hh} \sim \tau_i$ 
the temperature dependence of the resistivity is strongly affected by
hole-hole scattering and becomes 
(for $T<T_F$) $\rho_{T}=\rho_0 +
a T + b T^2$, where $\rho_0=\rho(T=0)$ and $a,b$ are positive
density dependent constants.
Our numerical calculation shows $a>b$ when $\tau_i>\tau_{hh}$
and $a<b$ when $\tau_i<\tau_{hh}$.

\begin{figure}
\epsfysize=2.3in
\centerline{\epsffile{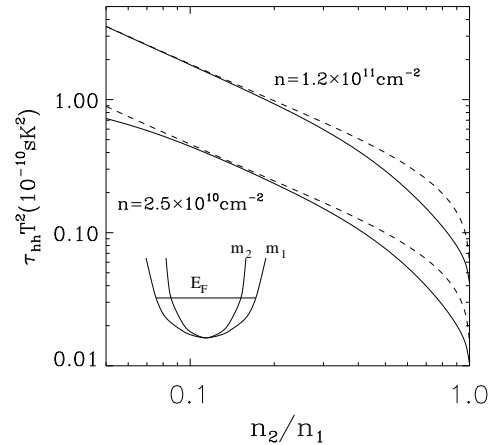} }  
\vspace{0.2cm}
\caption{
Calculated hole-hole scattering time $\tau_{hh}$ 
as a function of the density ratio 
$r=n_2/n_1$
for fixed total densities $n=1.2 \times 10^{11}cm^{-2}$,
$n=2.5 \times 10^{10} cm^{-2}$ at $T=0.1K$.
Solid (dashed) lines are the results from the screened 
Coulomb potential within the RPA (TFA). 
Inset shows the spin-split energy subbands.}
\end{figure}

In Fig. 1 we show the calculated hole-hole scattering time 
as a function of the density ratio $r=n_2/n_1$ for fixed total densities,
$n=n_1+n_2=2.5\times10^{10}cm^{-2}$, $1.2 \times 10^{11}cm^{-2}$.
Inset shows schematically the spin-split energy subbands with different 
effective masses.
We use a fixed effective mass $m_2=0.3m_e$ throughout this paper 
and create unequal population of the subbands (spin-splitting)
by tuning $m_1 > m_2$.
(Since Eq. (\ref{tauhh}) is symmetric with respect 
to the interchange of indices 1 and 2 we show the results
for $n_2/n_1 <1$).
Fig. 1 shows that the scattering time of RPA
is different from that of TFA when the density difference is small. 
Since most experimental situations \cite{papa,ham} 
correspond to $0.5<n_2/n_1 <1$ the use of the proper screening function
(i.e. RPA) is crucial (and long wavelength delta function scattering 
of TFA is inadequate), especially at low densities where 
the spin-splitting is very small. 
Fig. 1 also shows
that at a given density and temperature the hole-hole scattering time 
increases as the population difference of the two bands increases,
and the inverse scattering time $\tau_{hh}^{-1}$ diverges logarithmically as 
$r \rightarrow 1$. (This follows analytically from Eq. (\ref{tauhh}) 
also.)
The scattering time has strong density dependence
at a fixed temperature:
the scattering time decreases as the density decreases.

The effect of changing the confining potential 
is shown in Figs. 2 and 3, where we show the calculated fractional
change in the resistivity 
$\rho(T)/\rho(0)$ as a function of 
temperature. In Fig. 2 the results for a high density sample 
$n=1.2\times10^{11}cm^{-2}$ are given.
The density ratio $r$
decides the strength of the spin-splitting. (Smaller $r$ indicates
larger spin-splitting.)
For $n_1=n_2$ the fractional change in the resistivity 
shows a linear $T$ dependence and is determined
entirely by the mobilities of each band. 
In this case we have exactly
the same results as the single band model with $g_s=2$ and 
$n=n_1+n_2$.
When we tune 
\begin{figure}
\epsfysize=2.2in
\centerline{\epsffile{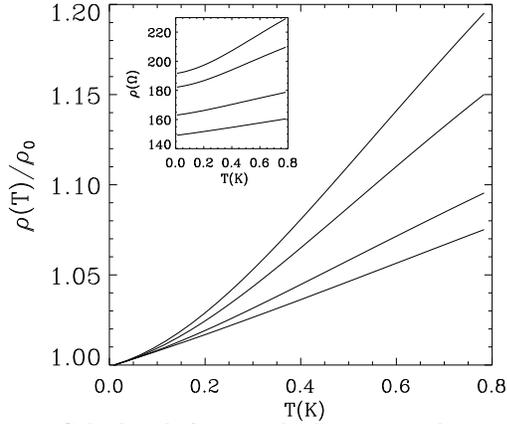} }  
\caption{Calculated fractional change in the resistivity 
$\rho(T)/\rho(0)$ as a function of 
for $n=1.2\times10^{11}cm^{-2}$. 
Various curves correspond to
$r= 1.0, \; 0.83,\; 0.67, \; 0.59$ (from bottom to top).
The temperature dependent resistivity for a given density is shown 
in insets.}
\end{figure}
\noindent
$m_1$ (keeping $m_2$ constant) the two subbands have
unequal densities, $\Delta n = n_1-n_2 =n(1-r)/(1+r)$.
As $r$ decreases the relative magnitude of the resistivity increases 
with temperature. Thus, 
the strength of the ``metallicity''
is enhanced when the two spin-split 
subbands have more unequal populations (i.e., larger spin splitting).
This shows that hole-hole carrier scattering 
enhances the metallic behavior, particularly at high densities.
(We get similar results for other high density samples.)

In Fig. 3 we show the results for a low density system, $n=2.5\times
10^{10} cm^{-2}$. 
At this density experiment \cite{papa} shows that the lowest temperature 
resistivity, $\rho_0$,
is an order of magnitude larger than that at $n=1.2\times10^{11}cm^{-2}$.
In addition, $\rho_0$ increases by about $100\%$
with increasing back gate bias 
even though the total carrier density is constant. (Note that in the 
high density samples $\rho_0$ increases by $\le 40\%$
with the same increase of the gate bias.)
We believe that in low density systems the effective disorder 
becomes larger when the confining potential 
becomes asymmetric (due to
the back gate bias). The important role of disorder in the 2D metallic
behavior under a back gate bias has also recently been emphasized in Si 
inversion layer experiments \cite{Lewalle}. 
To incorporate \cite{stern,sds} 
the disorder induced collisional broadening corrections 
we introduce the Dingle temperature $T_D$
in the screening function.
The pure RPA ($T_D=0$) case  
completely neglects collisional broadening effects on screening.
The effect of 
the Dingle temperature is to
increase the zero temperature resistivity 
(by suppressing screening for $T \ll T_D$)
as the splitting becomes large.
Fig. 3 shows that the 
temperature dependence of the resistivity 
is strongly enhanced when we consider a fixed Dingle temperature
and hole-hole scattering. Without hole-hole scattering (and with a fixed
Dingle temperature) the spin-split system still has a stronger
metallic behavior than the corresponding one subband
spin degenerate system ($r=1$), but
\begin{figure}
\epsfysize=2.1in
\centerline{\epsffile{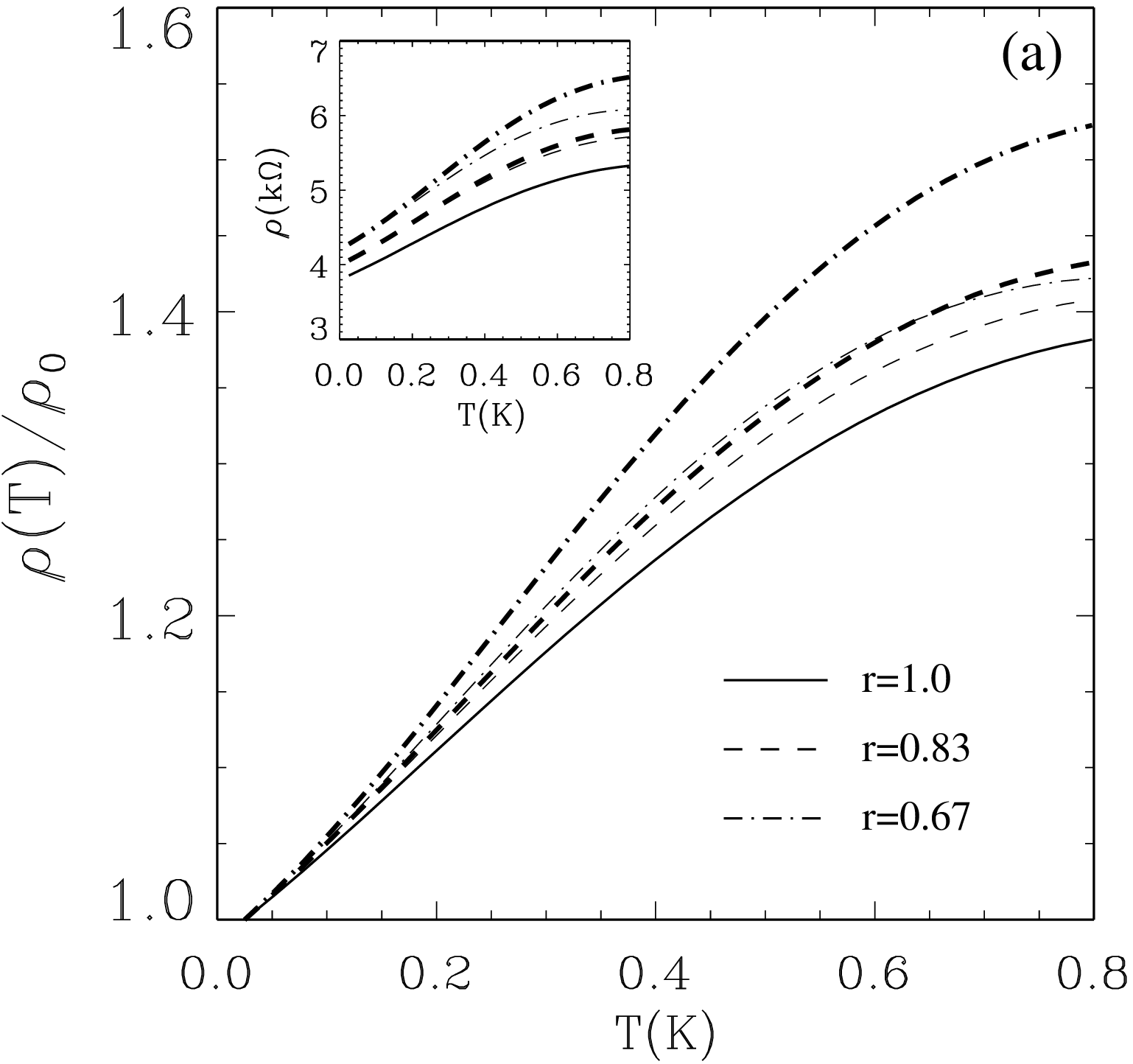} }  
\vspace{0.3cm}
\epsfysize=2.1in
\centerline{\epsffile{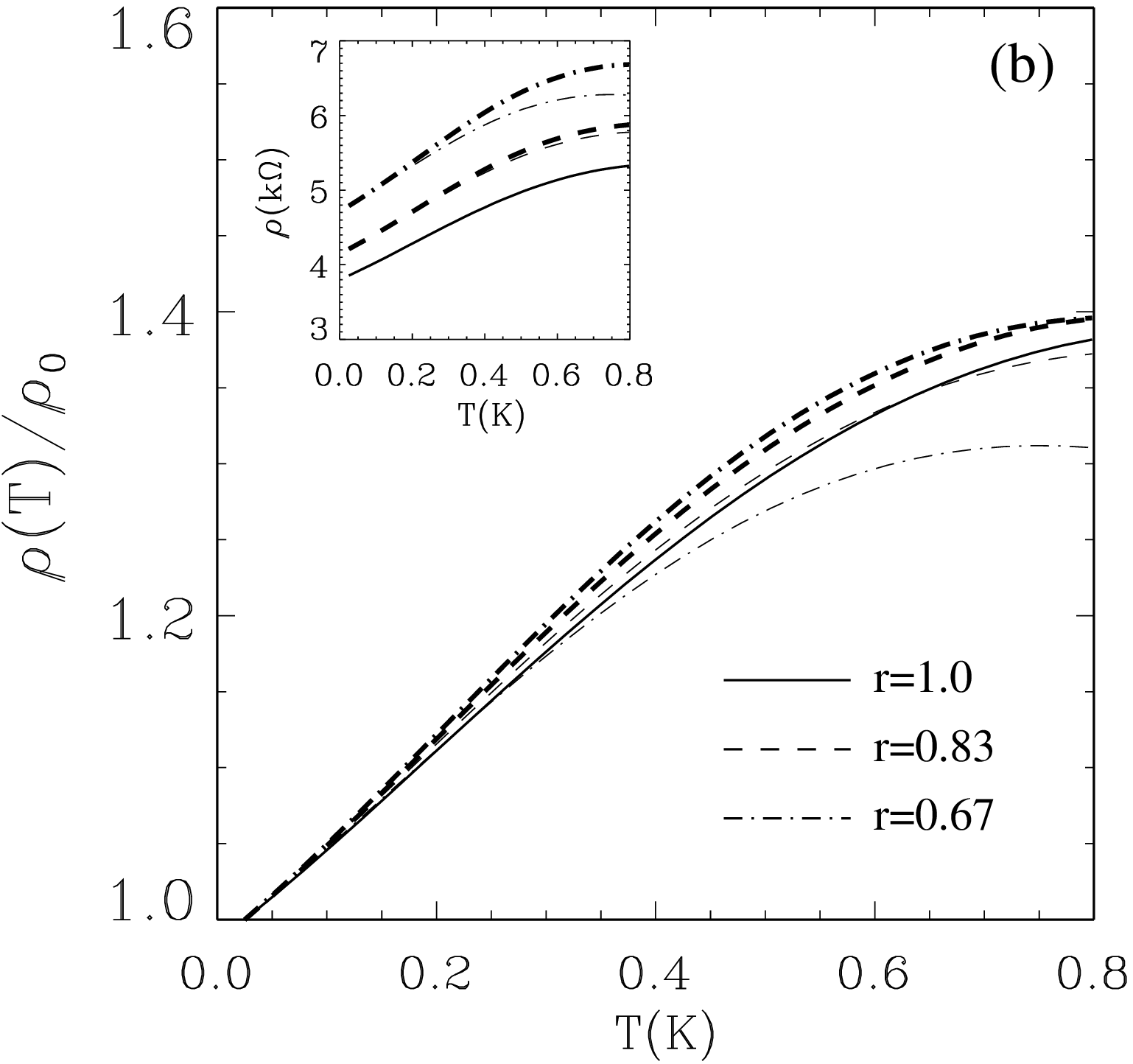} }  
\caption{
Same as in Fig. 2 for a low density systems with
$n=2.5\times 10^{10} cm^{-2}$. 
Solid lines are for the equal density case ($n_1=n_2$).
Thin (thick) lines are the results without (with) hole-hole scattering.
In (a) we use a fixed Dingle temperature ($T_D=0.1T_F$) 
for different $r$,
and in (b) we use different Dingle temperatures;
$T_D=0.1, \; 0.15, \; 0.2T_F$ for $r=1.0, \; 0.83,\; 0.67$, 
respectively. The insets show $\rho(T)$ for the various case in the main 
figures.}
\end{figure}
\noindent
the strength of the temperature dependence is reduced 
compared to the results with hole-hole scattering [Fig. 3(a)].
When we vary the Dingle temperature (i.e., increase the level broadening) 
the metallic behavior without hole-hole scattering
decreases as the spin-splitting increases (Fig. 3(b)),
which is precisely what is found experimentally \cite{papa}. 
Thus, without introducing the hole-hole scattering 
we can obtain qualitative agreement with experimental results for 
lower density samples including only disorder and
temperature dependent screening.

It is unclear whether our model including only 
spin-splitting induced hole-hole scattering and
screening induced temperature dependent disorder scattering 
can entirely quantitatively
describe the metallic behavior observed in low density 2D hole 
systems. 
But it 
is clear that our zeroth order model should be an essential part of 
any theory in this subject. We have established beyond any reasonable 
doubt that both spin-splitting induced hole-hole scattering and 
temperature dependent screening play important roles in the 
observed ``metallicity'' of 2D p-GaAs systems.

This work is supported by the U.S.-ONR the NSF, and DARPA.


\begin{thebibliography}{99}


\bibitem{one} E. Abrahams {\it et al.}, 
Rev. Mod. Phys. {\bf 73}, 251 (2001); B. L. Altshuler {\it et al.}, 
Physica E {\bf 9}, 209 (2001).
  
\bibitem{hwang} S. Das Sarma and E. H. Hwang, \prl {\bf 83}, 164 (1999);
\prb {\bf 61}, R7838 (2000); V. Senz {\it et al.}, cond-mat/0107369. 

\bibitem{stern}
F. Stern and S. Das Sarma, Solid State Electron. {\bf 28}, 158 (1985);
A. Gold and V. T. Dolgopolov, \prb {\bf 33}, 1076 (1986).

\bibitem{sds}S. Das Sarma, 
\prb {\bf 33}, 5401 (1986).

\bibitem{papa} S. J. Papadakis {\it et al}., Science {\bf 283},
2056 (1999); \prb {\bf 62}, 15375 (2000).

\bibitem{ham} A. R. Hamilton {\it et al.}, \prl {\bf 87}, 126802 (2001).

\bibitem{murzin} S. S. Murzin {\it et al}., Pis'ma Zh. Eksp. Teor. Fiz. 
{\bf 67}, 101 (1998) [JETP Lett. {\bf 67}, 113 (1998)]; 
Y. Yaish {\it et al}., \prl {\bf 84} 4954 (2000).


\bibitem{eisen} J. P. Eisenstein {\it et al}., \prl {\bf 53}, 2579 (1984);
Yu. A. Bychkov and E. I. Rashba,  Pis'ma Zh. Eksp. Teor. Fiz. 
{\bf 39}, 66 (1984) [JETP Lett. {\bf 39}, 78 (1984)]


\bibitem{gant} V. F. Gantmakher and Y. B. Levinson, Zh. Eksp. Teor. Fiz. 
{\bf 74}, 261 (1978) [Sov. Phys. JETP {\bf 47}, 133 (1978)];
C. A. Kukkonen and P. F. Maldaldague, \prl {\bf 37}, 782 (1976).


\bibitem{appel} J. Appel and A. W. Overhauser, \prb {\bf 18}, 758 (1978).


\bibitem{ando} S. Das Sarma {\it et al.}, \prb {\bf 19}, 6397 (1979).

\bibitem{siggia} E. D. Siggia and P. C. Kwok, \prb {\bf 2}, 1024 (1970),



\bibitem{Lewalle}A. Lewalle {\it et al.}, cond-mat/0108244.



\end{thebibliography}
\end{document}